\documentclass[nofootinbib,preprint,prb,preprintnumbers,superscriptaddress]{revtex4}
\usepackage{color}
\usepackage[latin9]{inputenc}
\usepackage{dcolumn}% Align table columns on decimal point
\usepackage{float}
\usepackage{natbib} 
\usepackage{graphicx}
\begin{document}

\title{A comprehensive scenario of the thermodynamic anomalies of water using the TIP4P/2005 model
\footnote{Accepted for publication in The Journal of Chemical Physics, \\
tentative reference: vol 145 issue 6 number 025630 \\ \\}
}

\author{Miguel A. Gonz\'alez}
\affiliation{Depto. Qu\'{\i}mica F\'{\i}sica I, Fac. Ciencias Qu\'{\i}micas,
Universidad Complutense de Madrid, 28040 Madrid, Spain}
\affiliation{Department of Chemistry, Imperial College London, London SW7 2AZ, United Kingdom}
\author{Chantal Valeriani}
\affiliation{Depto. Qu\'{\i}mica F\'{\i}sica I, Fac. Ciencias Qu\'{\i}micas,
Universidad Complutense de Madrid, 28040 Madrid, Spain}
\affiliation{Depto. F\'{\i}sica Aplicada I, Fac. Ciencias F\'{\i}sicas, 
Universidad Complutense de Madrid, 28040 Madrid, Spain}
\author{Fr\'ed\'eric Caupin}
\affiliation{Institut Lumi\`ere Mati\`ere, UMR5306 Universit\'e Claude Bernard Lyon 1-CNRS,
Universit\'e de Lyon, 69622 Villeurbanne Cedex, France}
\author{Jos\'e L. F. Abascal
}

\affiliation{Depto. Qu\'{\i}mica F\'{\i}sica I, Fac. Ciencias Qu\'{\i}micas,
Universidad Complutense de Madrid, 28040 Madrid, Spain}

%\date{\today}
%
\begin{abstract}
The striking behavior of water has deserved it to be referred to as an
``anomalous" liquid. The water anomalies are greatly amplified in metastable
(supercooled and/or stretched) regions.
This makes difficult a complete experimental description since, beyond certain
limits, the metastable phase necessarily transforms into the stable one.
Theoretical interpretation of the water anomalies could then be based on
simulation results of well validated water models. But the analysis of the
simulations has not yet reached a consensus. In particular, one of the
most popular theoretical scenarios ---involving the existence of a
liquid-liquid critical point (LLCP)--- is disputed by several authors.
In this work we propose to use a number of exact thermodynamic relations which
may shed light on this issue.
Interestingly, these relations may be tested in a region of the phase diagram
which is \textit{outside} the LLCP thus avoiding the problems associated to the
coexistence region.
The central property connected to other water anomalies is the locus of
temperatures at which the density along isobars attain a maximum (TMD line) or
a minimum (TmD).
We have performed computer simulations to evaluate the TMD and TmD for a
successful water model, namely TIP4P/2005.
We have also evaluated the vapor-liquid spinodal in the region of large
negative pressures.
The shape of these curves and their connection to the extrema of some response
functions, in particular the isothermal compressibility and heat capacity at
constant pressure, provide a very useful information which may help to
elucidate the validity of the theoretical proposals.
In this way we are able to present for the first time a comprehensive scenario
of the thermodynamic water anomalies for TIP4P/2005 and their relation to the
vapor-liquid spinodal.
The overall picture shows a remarkable similarity with the corresponding one
for the ST2 water model, for which the existence of a LLCP has been
demonstrated in recent years.
It also provides a hint as to where the long-sought for extrema in response
functions might become accessible to experiments.
\end{abstract}

\maketitle

\section{Introduction}

The physical properties of water at ambient conditions are markedly different
from those of other liquids. It is widely known that the density of liquid
water at a fixed pressure exhibits a maximum at the so-called temperature of
maximum density (TMD). In particular, at atmospheric pressure, the TMD is
approximately 4 $^{\circ}$C. Below this temperature the water's expansivity,
$\alpha$, is negative in striking contrast with ``normal" liquids where
$\alpha$ is always positive. Other thermodynamic response functions, such as
the isothermal compressibility, $\kappa_T$, or the isobaric heat capacity $C_p$
also show an unusual behavior.\cite{debenedetti03} On supercooling, these
anomalies are enhanced. In particular, the isothermal compressibility seems to
diverge at 228~K.\cite{speedy76} 

Several scenarios have been proposed to account for the water 
anomalies and their magnification at temperatures below the melting
point.\cite{speedy82,poole92,sastry96,angell08}
In the stability limit conjecture (SLC),\cite{speedy82} the increase of the
response functions in the supercooled region is ascribed to a continuous
retracing line of instability that delimits the supercooled and stretched
metastable states.
The second critical point scenario assumes that there is a liquid-liquid
coexistence terminating at a critical point (LLCP).\cite{poole92}
The response functions would reach a peak and converge towards the Widom line
(the line of the maxima of the correlation length) emanating from the
LLCP.\cite{sciortino97,xu05}
Finally, it has been shown that thermodynamic consistency explains the
existence of peaks in the response functions as a mere consequence of the
presence of density anomalies. In this case there is no singular behavior,
hence the name of singularity-free (SF) scenario.\cite{sastry96} Notice that
the SF interpretation can also be seen as the second critical-point hypothesis
with the LLCP occurring at zero temperature.

The experimental testing of these scenarios  is extremely difficult, if not
impossible, because of the difficulties to access    the ``no man's land" region
that lies below the temperature at which water spontaneously freezes.
Crystallization may be inhibited by confining water in nanosized samples, but
it is unclear whether surface effects could influence the outcome of the
experiments.\cite{zanotti05,mallamace07a,mallamace08}
Different experimental approaches have been proposed to circumvent the problems
associated with the bulk water no man's land\cite{mishima98,mishima98b,bellissent-funel98,soper00,mishima02,souda06,banerjee09,taschin13,pallares14,sellberg14,seidl15} (see also a recent review\cite{caupin15} on this topic).
They have provided very important information on the behavior of water at
extreme conditions (deep supercooling and/or high negative pressures). Although
the experiments seem to be consistent and support the appearance of a
liquid-liquid transition, their interpretation is not conclusive.

Theoretical work has shown that the slope of the TMD loci determines the
behavior of the thermodynamic response functions.
Exact thermodynamic relations relate the shape of the TMD to that of other
water anomalies.
It is well known that the TMD of liquid water is a negatively sloped function
in a p-T diagram. If the negative slope would extend to large negative
pressures (SLC scenario), it would meet the vapor-liquid spinodal.
In such a case, thermodynamic consistency requires\cite{debenedetti86} that the
spinodal would retrace at the intersection point.
On the other hand, the TMD could change its slope leading to a nose-shaped
function (LLCP and SF scenarios).
It has been demonstrated that a positively sloped TMD line cannot cross a
positively sloped spinodal in a thermodynamically consistent phase
diagram and that the turning point of the TMD must
intersect the locus of isothermal compressibility extrema.\cite{sastry96}
In summary, the study of the TMD of stretched water may give insight to    the
validity of the hypothesis proposed to explain the water anomalies.
Recent measurements of the speed of sound\cite{pallares14} have enabled to
extend considerably our knowledge of the TMD in the region of large negative
pressures.\cite{pallares16}
These results indicate that the slope of the TMD becomes increasingly more
negative as the pressure decreases and strongly suggest that the experimental
TMD is about to reach a retracing point. Unfortunately, bubble nucleation
prevents carrying this study further.

Given the experimental difficulties, it is clear that molecular simulation may
be an alternative for our purpose. Most of the computer simulation
studies using realistic water models seem to support the existence of the
liquid--liquid separation. The seminal work of Poole et al.\cite{poole92} 
focused on the ST2 water model.\cite{stillinger74} Some of the features of
the ST2 model allow a thorough investigation of the supercooled region. Because
of this, the model has been widely used in the study of the liquid-liquid
phase transition. Most simulations using ST2%
\cite{poole93,sciortino97,xu05,brovchenko05,liu09,sciortino11,liu12,kesselring12,poole13,palmer13,palmer14,yagasaki14,palmer16}
seem to have unambiguously demonstrated the existence of a LLCP although
this interpretation has been challenged by Limmer and 
Chandler.\cite{limmer11,chandler16}
But the advantages of the model for the study of the supercooled region (among
them, a high value for the TMD) are closely related to its major drawback: ST2
is known to produce an over-structured liquid compared to real water. Thus,
there is no compelling evidence that the behavior of metastable water can be
described by ST2.\cite{liu09}

Alternative successful water models can indeed be found in the literature
though they are not free of objections. SPC/E\cite{berendsen87} is a widely used
model showing excellent predictions for a number of water properties in the
liquid region.\cite{vega11} However, its bad performance in locating
the temperature of maximum density (TMD), the melting temperature, T$_{m}$, and
the isothermal compressibility minimum\cite{pi09}, seem to discourage its use
to investigate the supercooled region.
Since TIP5P\cite{mahoney00,rick04} provides very good estimates of both the
TMD and T$_{m}$, it has been used in simulation studies attempting to disclose
the behavior of metastable liquid water.%
\cite{yamada02,paschek05,brovchenko05,xu05,brovchenko05,yagasaki14}
However, the excellent performance of the model at ambient conditions is not
preserved when one moves away of this region. This failure is particularly
serious because it is a signal that the results for the response functions
cannot be satisfactory.\cite{vega11} Moreover, TIP5P gives a very poor estimate
of the density of hexagonal ice and, hence, of the density and other properties
of a possible low density phase in a liquid-liquid coexistence.

It has been demonstrated \cite{vega05a,abascal07a,abascal07c} that the
TIP4P geometry is more appropriate than that of three-site models ---such as
SPC--- or five-site models ---like TIP5P--- to account for the TMD and the
liquid-solid equilibrium of water. These studies followed an increasing
interest in re-parametrized TIP4P models.\cite{horn04,abascal05a,abascal05b}
Among these, TIP4P/2005\cite{abascal05b} seems to produce a better overall
agreement with experiment for a large number of properties of water in
condensed states.\cite{vega09,vega11}
Moreover, TIP4P/2005 results are quite accurate for properties relevant to the
study of metastable water, namely water anomalies\cite{pi09} and equation of
state of supercooled water.\cite{abascal11}
Finally, the model gives a quantitative account of recent measurements of the
speed of sound of doubly metastable (supercooled and stretched)
water.\cite{pallares14}

From the above arguments it seems then that TIP4P/2005 is the ideal candidate
for the study of the water anomalies in the supercooled region. It may come as a
 surprise that only a reduced number of works have been devoted to this
issue,\cite{abascal10,sumi13,limmer13,overduin13,yagasaki14,bresme14,russo14,overduin15,singh16}
probably because the model has also some limitations mainly derived from the
large structural relaxation times at deeply supercooled states.
Abascal and Vega\cite{abascal10} proposed that the model exhibits a LLCP at 
193~K and reported a case of a liquid-liquid separation (low- and high-density)
below the second critical point. 
Even though Overduin and Patey\cite{overduin13} argued that longer
simulations (8~$\mu$s for 500 molecules, instead of 400~ns) were necessary to
obtain well converged density distributions at those conditions, 
a number of authors\cite{sumi13,yagasaki14,bresme14,russo14} confirmed 
the rest of the results presented in Ref.~\onlinecite{abascal10}.
The study of Overduin and Patey does not essentially contradict the
results of Abascal and Vega if the suggested LLCP of TIP4P/2005 would be
slightly shifted towards lower temperatures.
This is in line with the critical temperature reported for this model by Sumi
and Sekino\cite{sumi13} (182~K) and Yagasaki {\it et al.}\cite{yagasaki14}
(185~K). Interestingly, a two-structure equation of state consistent with the
presence of a LLCP provides a very similar critical
temperature.\cite{bresme14,singh16} Therefore, it would be of great
interest to perform a simulation study using advanced sampling methods to
unambiguously check the existence of a LLCP for this model (similar to that
successfully accomplished for ST2\cite{palmer14}).
However, the work of Overduin and Patey clearly indicates that such study would
be extremely costly in computer time.

In this work we propose to circumvent the question of the existence of a LLCP
and focus on the related issue of the shape of the TMD and its relation to
other water anomalies. As shown above, the study not only involves calculations
in the supercooled and/or stretched region \textit{outside} the proposed
critical region, but it may also provide a complete perspective of the scenario
of water anomalies. Although the study is highly demanding in computer
resources, it is still affordable.
We also note that, regardless of the nature of the phase diagram of TIP4P/2005
liquid at low temperature (real or virtual critical point, \ldots), the region
we consider in the present paper includes the one relevant for experiments on
bulk water. It can therefore serve as a guide for future measurements.

\section{Methods}

All simulations (except those intended for the calculation of the vapor-liquid
spinodal) have been performed with 4\,000 TIP4P/2005 water molecules in the
isothermal-isobaric $NpT$ ensemble using the Molecular Dynamics package GROMACS
4.6\cite{spoel05,hess08} with a 2~fs timestep.
Long range electrostatic interactions have been evaluated with the smooth
Particle Mesh Ewald method.\cite{essmann95}
The geometry of the water molecules has been enforced using \textit{ad hoc}
constraints, in particular, the LINCS algorithm.\cite{hess97,hess08b}
To keep the temperature and pressure constant, the Nos\'e-Hoover
thermostat\cite{nose84,hoover85} and an isotropic Parrinello-Rahman barostat
have been applied\cite{parrinello81} with 2~ps relaxation times.

Most of our calculations intended to evaluate the extrema of thermodynamic
properties, and we have adapted our strategy according to this goal.
First, we calculated the desired property at regular intervals along
isotherms/isobars to provide a rough estimate of the position of the maximum or
minimum.
Then we ran additional points to precisely locate it.
We monitored the uncertainties along the simulation and extended the runs until the
differences between the consecutive points were larger than the statistical
uncertainty.
Thus, the required simulation times varied widely for the different properties
and state points.
Since most of the calculations correspond to regions where the relaxation of
the system is quite slow, the length of the simulations is often of the order
of a few hundreds of ns, reaching 1.3~$\mu$s for the longest run.
Despite the careful monitoring of the runs to save computer resources, the
required simulation times together with the use of a relatively large system
size implies an important computational effort (equivalent to more than 300\,000
hours of 2.6 GHz Xeon cores) that has been achieved by means of a GPU-based
supercomputer.

The uncertainty on each measurement has been calculated using a method proposed
by Hess.\cite{hess02b}
The trajectory is divided in blocks, the average for each block is
calculated and the error is estimated as the standard deviation of the block
averages.
Also, an analytical block average curve is obtained by fitting the
autocorrelation between block averages to a sum of two exponentials.
In this way, the calculated uncertainties lead to an asymptotic curve only if
the trajectory is long enough so that the blocks are uncorrelated.
In summary, the procedure not only provides an estimate of the error but also
sheds light on the convergence of the trajectory.
An example of the application of the method is given as supplementary
material.\cite{epaps}

\section{Results}
\begin{center}
\begin{figure}[!ht]
\caption{Densities predicted by the TIP4P/2005 model compared to recent experimental data\cite{pallares16} for the 0.1~MPa, -40~MPa and -80~MPa isobars.}
\includegraphics*[clip,scale=0.6]{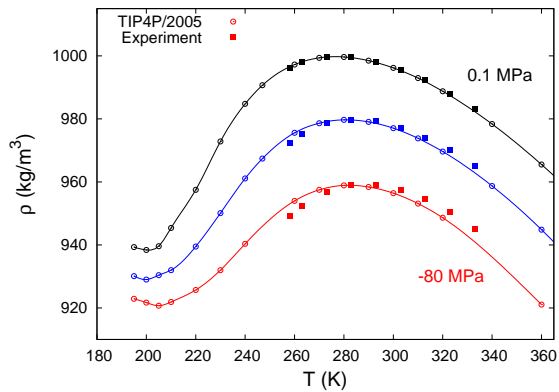}
\label{fig:dens_pallares}
\end{figure}
\end{center}

Although TIP4P/2005 provides quite acceptable results for the density of water
at positive pressures (also including the supercooled region\cite{abascal11}),
its performance at negative pressures has not been thoroughly assessed (see
however Refs.~\onlinecite{pallares14} and \onlinecite{pallares16}).
Very recently, an experimental equation of state for water down to -120~MPa
has been reported.\cite{pallares16}
This allows to check for the first time the predictions for the equation of
state in the large negative pressures region.
predictions 
The numerical values of the density along some isobars for the TIP4P/2005 model
are given as supplementary material.\cite{epaps}
Figure~\ref{fig:dens_pallares} shows that the agreement between simulation
results and experiment is excellent although the departures increase with
decreasing pressures.
As a consequence, the prediction for the TMD is slightly shifted, the difference
at -80~MPa being about 7 degrees.

\begin{center}
\begin{figure}[!ht]
\caption{Locus of density maxima (TMD, thick line) and minima (TmD, thin line).
Points A and B mark the turning point of the TMD curve and the point at which
the TMD and TmD lines meet, respectively.}
\includegraphics*[clip,scale=0.5]{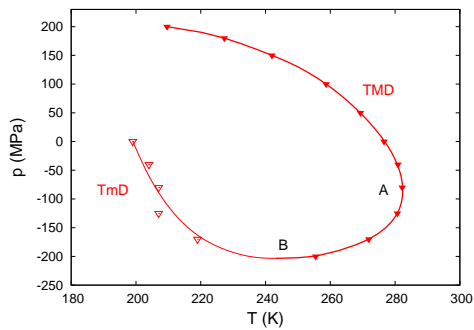}
\label{fig:dens}
\end{figure}
\end{center}

\begin{center}
\begin{figure}[!ht]
\caption{Detail of the densities for the -125~MPa (decreased by 0.8 Kg/m$^3$) and -170~MPa isobars showing the existence of a density minimum. The error bars correspond to the 90\% confidence interval calculated from the standard deviations obtained in the block procedure. Lines are a weighted fit of the data to a fourth order polynomial.}
\includegraphics*[clip,scale=0.6]{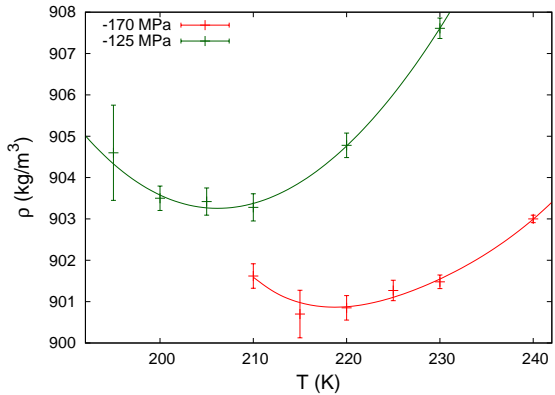}
\label{fig:dens-zoom}
\end{figure}
\end{center}

In the positive pressures region, the simulation data show a maximum density
for isobars up to a pressure of 200~MPa. In accordance with
experiment,\cite{angell76} at increasing pressures the TMD shifts to lower
temperatures.
At negative pressures the slope of the TMD becomes increasingly negative until
the curve retraces (point A in Fig~\ref{fig:dens}).
From the turning point to highly negative pressures the TMD has a positive slope. 
This result has already been reported in previous works for much smaller
samples.\cite{agarwal11,pallares16}
Our results for the larger system are very similar to the previous ones and
indicate that finite size effects in this region (if any) are quite small (see
supplemental material\cite{epaps}).
The largest (negative) pressure for which we have been able to calculate the
temperature of maximum density is -170~MPa.
Unfortunately, at -200~MPa the system cavitated for several runs using 4\,000 
water molecules.
However, using 500 molecules allowed us to perform short runs before the
system cavitated so it is possible an approximate calculation of the densities
and the approximate position of the TMD at this pressure.
The isobars at negative pressures also exhibit a temperature of minimum density, TmD.
However, in the positive pressures region, only the 0.1~MPa curve shows the
density minimum.
In all cases, the minimum is quite shallow and it is barely appreciable,
especially in the case of the -125~MPa and -170~MPa isobars.
Fig.~\ref{fig:dens-zoom} show a detail of these isobars clearly demonstrating
the existence of density minima.
Despite the great computational effort and the notable accuracy of the
density calculations, the uncertainty of the temperatures of minimum density is
about 5 degrees and the loci of the TmD produce a less smooth curve than that of
the TMD ones (see Fig.~\ref{fig:dens}).

The existence of a density minimum in real water has not been described
previously. Liu \textit{et al.}\cite{liu07} have reported a density minimum in
deeply supercooled confined deuterated water.
At ambient pressure, the minimum density occurs at 210~K with a value of
1041~kg/m$^3$.
Despite that it is difficult to know how the confinement affects the water
properties we may use this data as a rough guide of the behavior of bulk water.
As to TIP4P/2005, previous simulation results indicated the existence of the
density minimum\cite{pi09,russo14,sumi13} but the accuracy of the data did not
allow for a trustworthy estimation of its value.
The result of the present work at 0.1~MPa is $\rho$=938.1~kg/m$^3$ and is
located at 200$\pm$5~K. 
Assuming that the density of deuterated water is 1.106~times that of normal
water,\cite{liu07} we get $\rho$=1038~kg/m$^3$, close to the experimental
result in confined water.
 
The difference between the temperature of maximum and minimum density has a
peculiar behavior because it is larger near the retracing point of the TMD
and decreases at both higher and lower pressures.
At large negative pressures, the TMD and TmD lines converge asymptotically
(point B in the bottom panel of Fig.~\ref{fig:dens}).
Below this pressure, the density no longer exhibits maxima nor minima.

\begin{center}
\begin{figure}[!ht]
\caption{Top: Isothermal compressibility as a function of temperature along
isobars. Bottom: locus of $\kappa_T$ extrema together with TMD line.
Both curves intersect at the point A, the retracing point of the TMD.
The curves are cubic splines to guide the eye.}
\includegraphics*[clip,scale=0.27,angle=270]{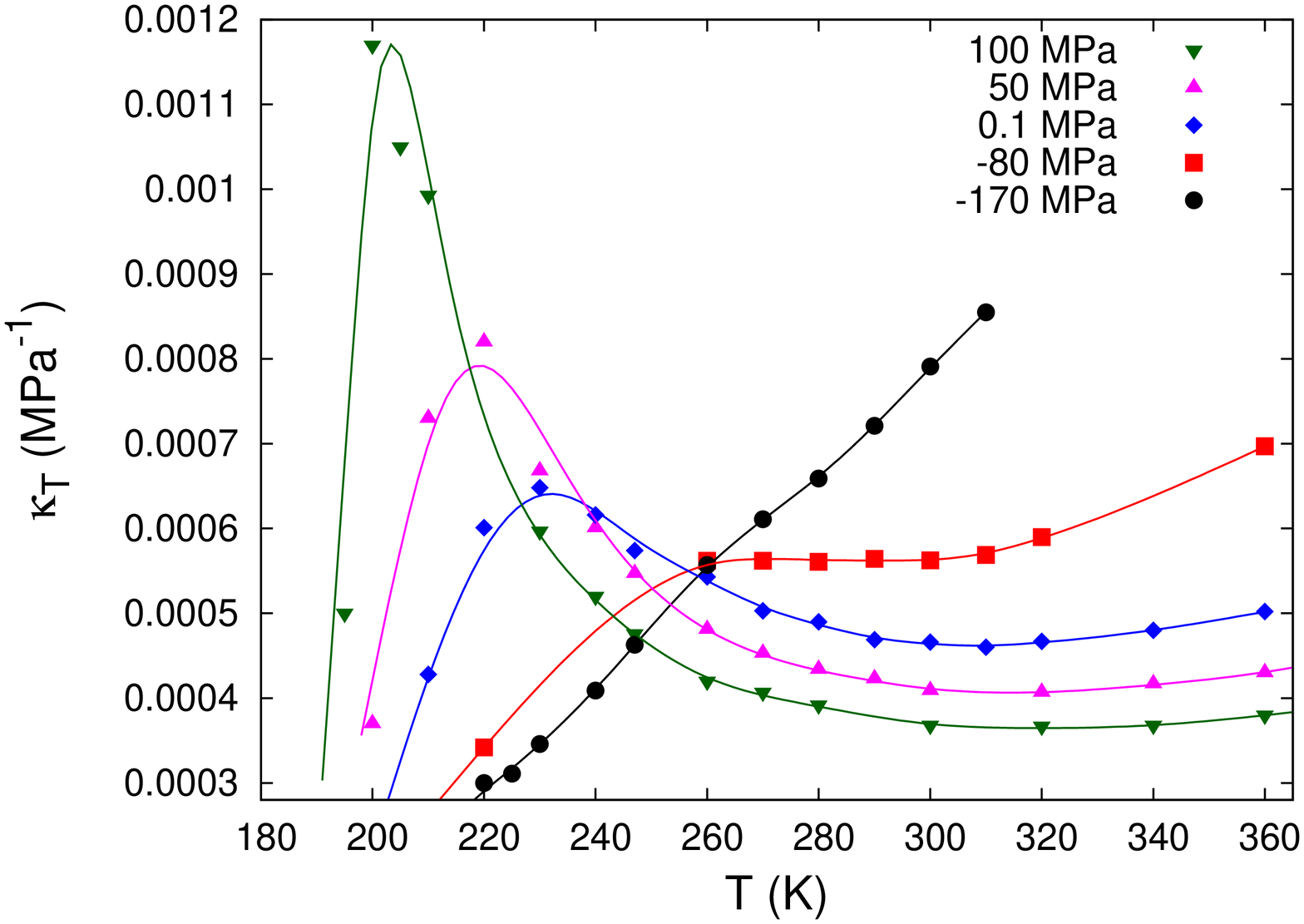}\\
\includegraphics*[clip,scale=0.27,angle=270]{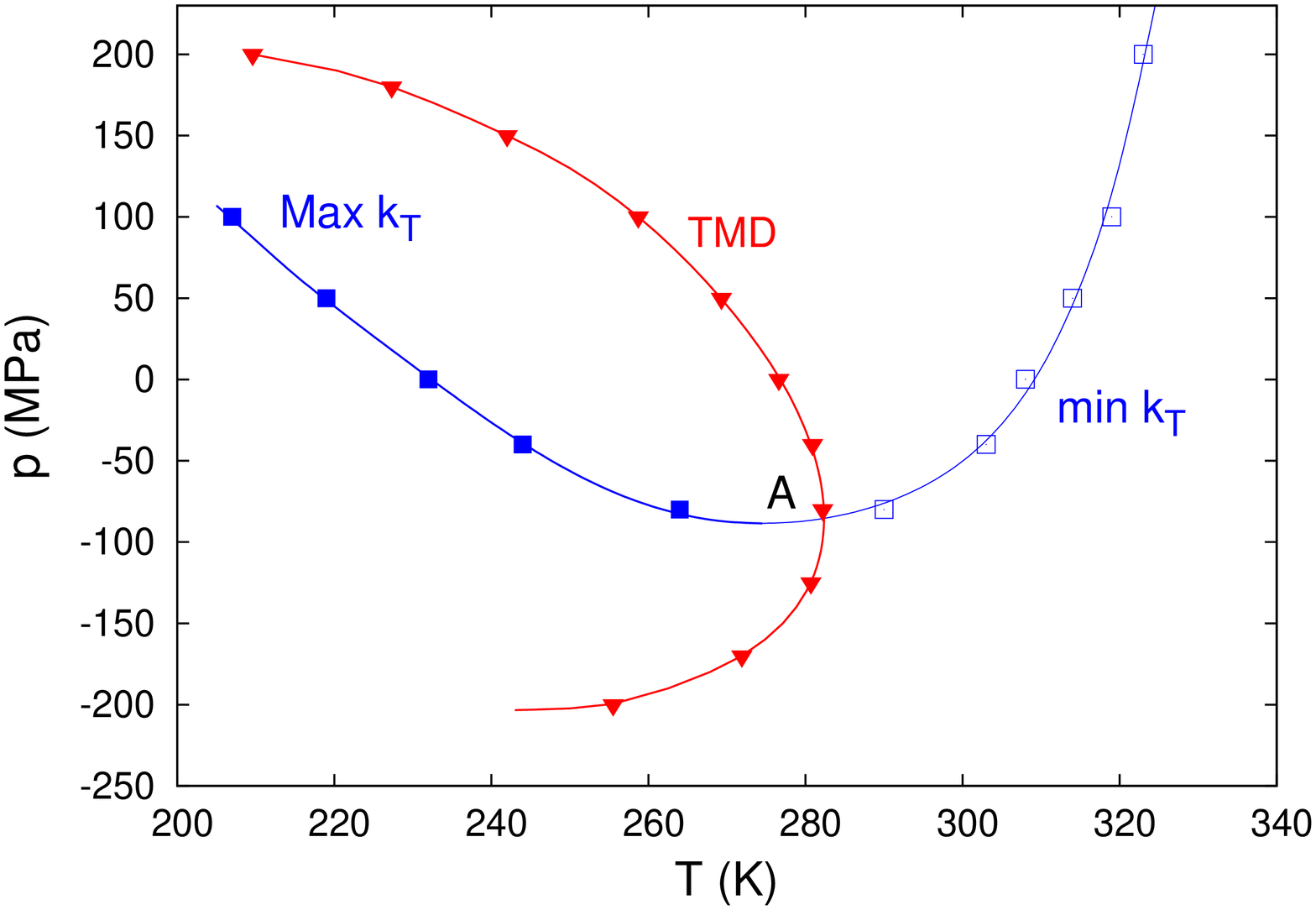}
\label{fig:kT}
\end{figure}
\end{center}

As commented in the introduction, at the retracing point, the TMD curve must be
crossed by the line joining the locus of isothermal compressibility extrema.
The upper panel of Figure~\ref{fig:kT} shows $\kappa_T$ as a function of
temperature for several isobars from -170~MPa to 100~MPa (for clarity only a
few of the simulated isobars are depicted).
For pressures higher than about -80~MPa the isobars show clearly the presence
of a maximum and a minimum.
At this pressure, the curve exhibits almost imperceptible extrema but, for a
slightly lower pressure, the maximum and minimum of $\kappa_T$ collapse into an
inflection point.
Then, at large negative pressures, $\kappa_T$ is a a monotonously increasing
function of temperature.
The locus of $\kappa_T$ extrema are plotted, in the p-T plane, in the lower
panel of Fig.~\ref{fig:kT} together with the TMD curve.
As expected, both lines cross at the turning point of the TMD (point A in
Fig.~\ref{fig:kT}).
Notice that the intersection point A lies near the pressure at which $\kappa_T$
becomes a monotonous function. 

\begin{center}
\begin{figure}[!ht]
\caption{Top: Heat capacity at constant pressure along isotherms.
Bottom: Locus of C$_p$ extrema together with the curve of density extrema.
Both lines intersect at point B separating the TMD and TmD.
The curves are a guide to the eye.}
\includegraphics*[clip,scale=0.5]{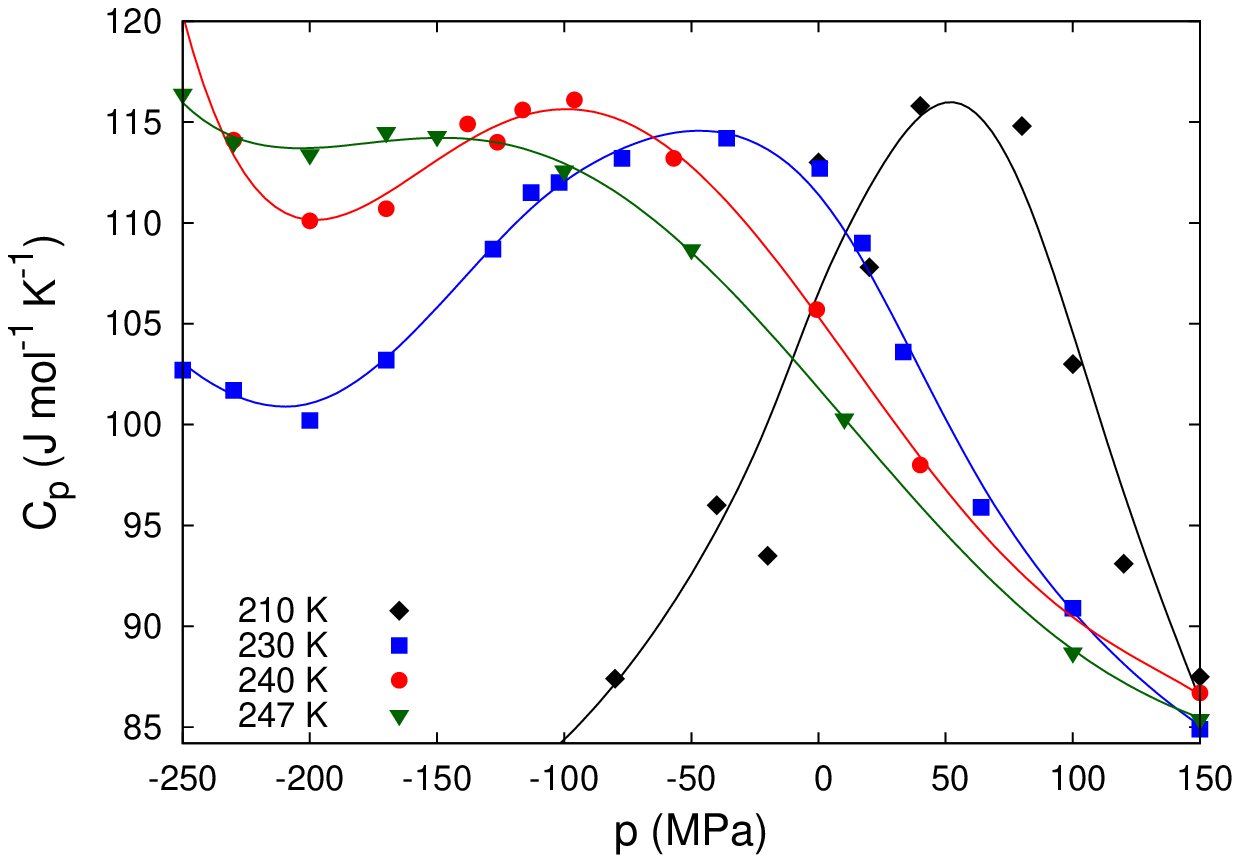}\\
\includegraphics*[clip,scale=0.5]{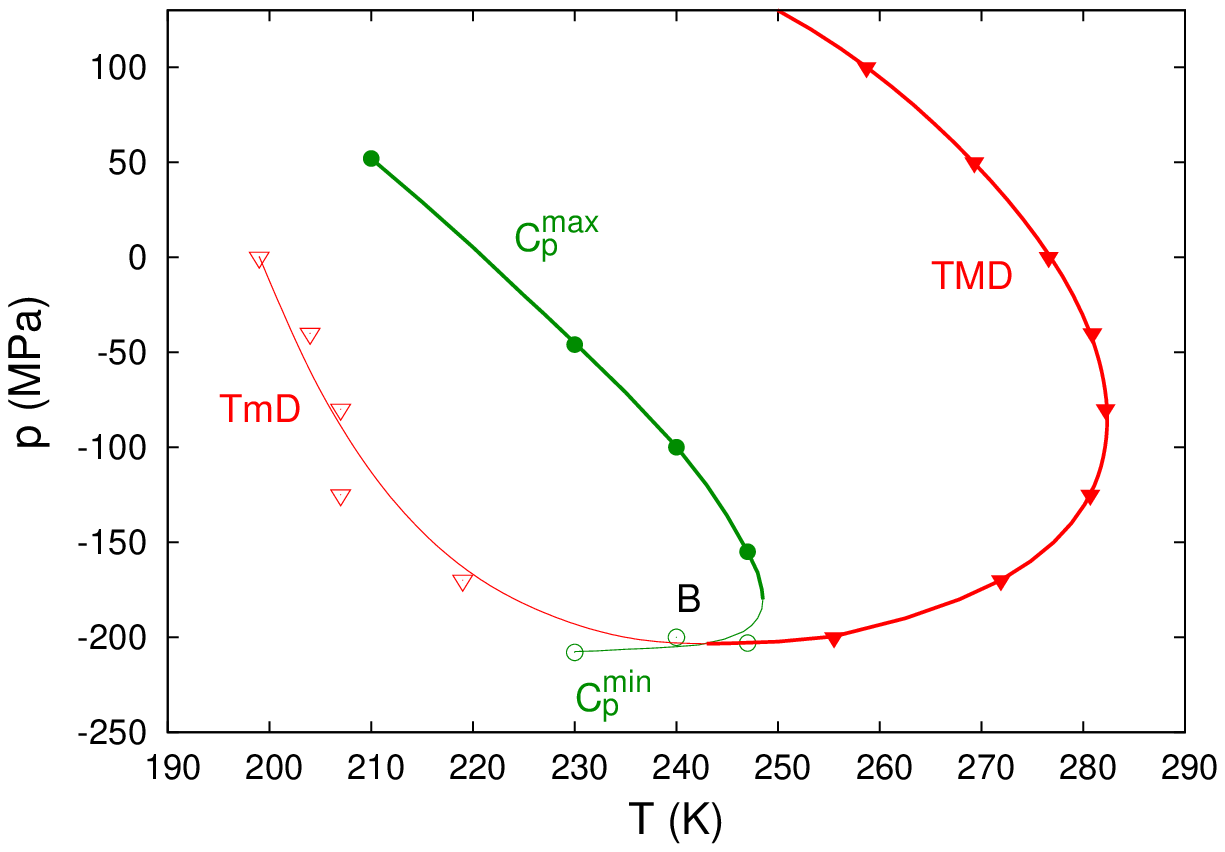}
\label{fig:cp}
\end{figure}
\end{center}

On the other hand, theoretical considerations\cite{poole05} indicate that the
locus of isobaric heat capacity extrema along isotherms separates
the TMD from the line of minimum densities, TmD. 
The simulation results for C$_p$ along isotherms are presented in
Figure~\ref{fig:cp} (upper panel).
The isotherms show both a maximum and a minimum.
The separation between these extrema becomes increasingly smaller as the
temperature increases.
Eventually, at a temperature slightly above 247~K, the maximum and minimum
converge to an inflection point and C$_p$ becomes a monotonous function.
The pressures at which the C$_p$ extrema occur for each isotherm are shown
in the lower panel of Fig.~\ref{fig:cp}.
In this figure we have also depicted  the locus of density extrema along
isobars. 
As shown by Poole \textit{et al.},\cite{poole05} the latter curve must have a
zero slope at the intersection point, a condition which is satisfactorily
fulfilled by our simulations (see point B in Fig.~\ref{fig:cp}).
The location of this point for TIP4P/2005 is about (243~K, -203.4~MPa).

\begin{center}
\begin{figure}[!ht]
\caption{Estimation of the vapor-liquid spinodal. Top: Pressure as a function
of the specific volume for several isotherms (asterisks indicate the position
of the minimum). Bottom: Specific volume as a function of temperature along the
vapor-liquid spinodal.}
\includegraphics*[clip,scale=0.6]{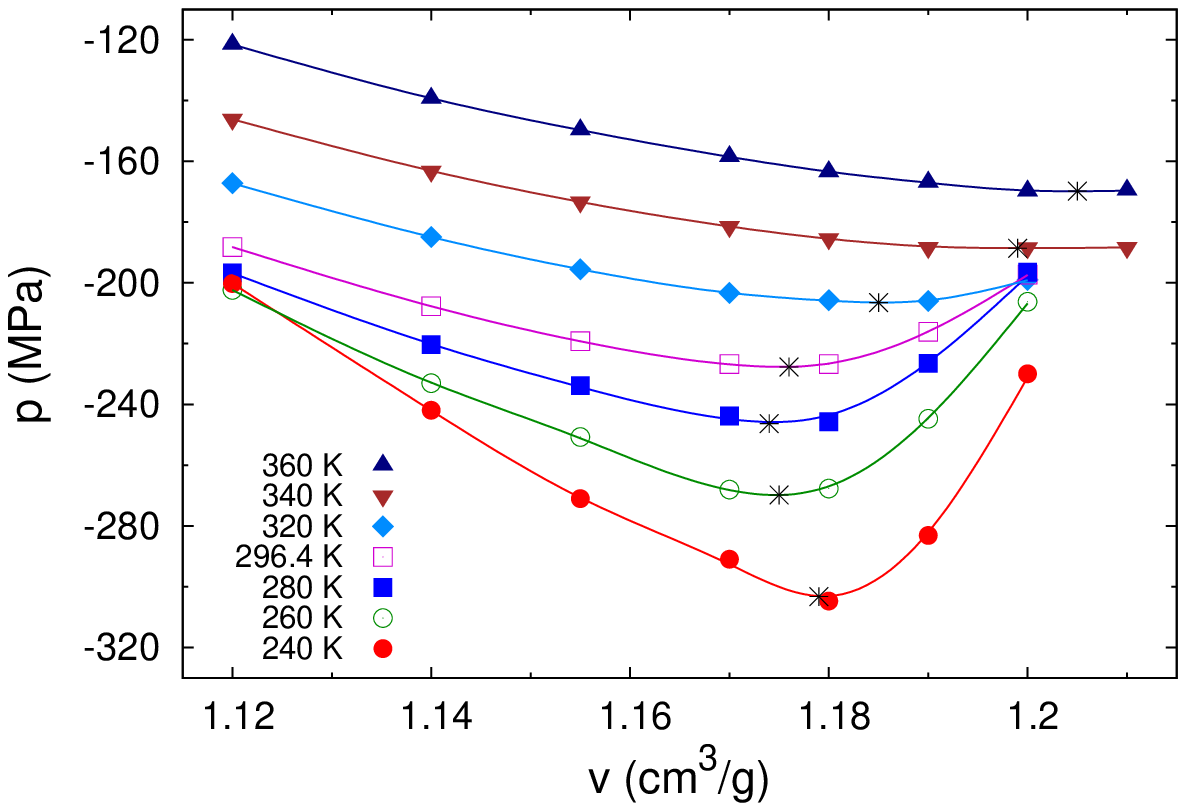} \\
\includegraphics*[clip,scale=0.6]{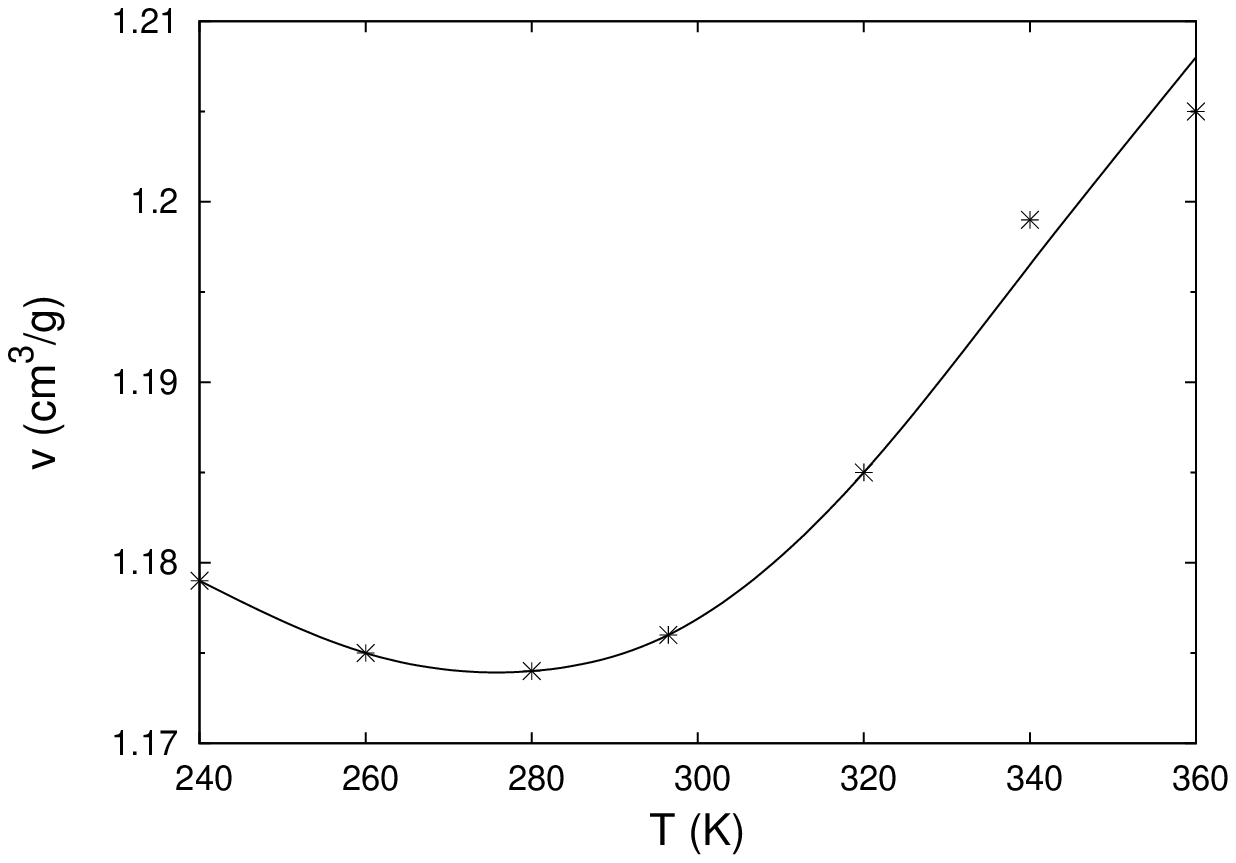}
\label{fig:spinodal}
\end{figure}
\end{center}

\begin{center}
\begin{figure}[!ht]
\caption{Comprehensive scenario relating thermodynamic water anomalies and the vapor-liquid spinodal. Dashed lines are an extension of the loci of $\kappa_T$ and $C_p$ maxima up to the proposed location of the LLCP.\cite{yagasaki14,singh16}}
\includegraphics*[clip,scale=0.8]{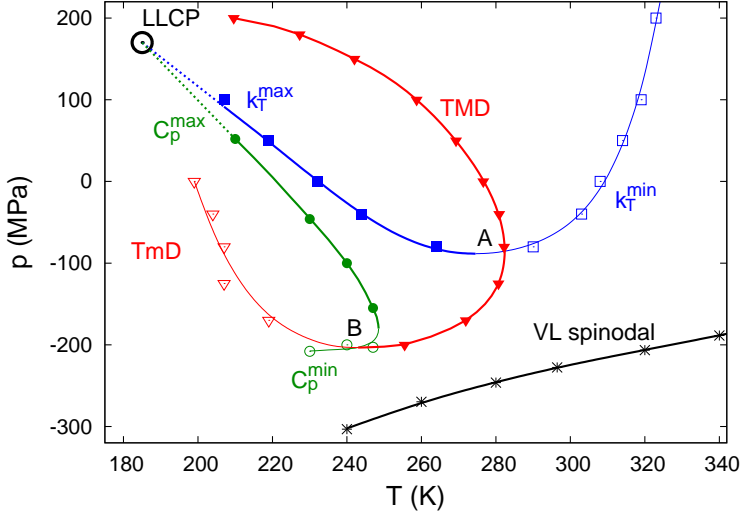}
\label{fig:full_scenario}
\end{figure}
\end{center}

It is clear at this point that the SLC conjecture is not valid for TIP4P/2005
and that the vapor-liquid (VL) spinodal should not meet the (retracing) TMD
line.
In the SF and LLCP scenarios both curves do not intersect.
It is then interesting to check whether this is fulfilled by our calculations.
We have tried to evaluate the VL spinodal by locating the zero slope
of the pressure-volume curves along isotherms.
The simulations were performed in the canonical (NVT) ensemble.
However, the system with 4\,000~molecules sometimes cavitated before providing
statistically significant results.
We were then forced to reduce the size of the system to 500~molecules.
For these samples, the probability of a cavitation event is almost one order
of magnitude smaller and it is then possible to obtain statistically
significant results.
It is well known that finite size effects may be important in this
region\cite{binder92} so our calculations must be seen as a first approximation
to the actual spinodal.
The corresponding pressure-specific volume isotherms are presented in
Figure~\ref{fig:spinodal} (top panel) from which we may extract the p-v-T
values of the VL spinodal. It is to be noticed that the specific volume along
the spinodal shows a non-monotonic dependence on both temperature and pressure
(bottom panel of Fig.~~\ref{fig:spinodal}).
As expected, the slope of the VL spinodal in the p-T plane is positive and do
not meet the TMD curve (Figure~\ref{fig:full_scenario}).

The results shown in the bottom panels of Figs.~\ref{fig:kT} and \ref{fig:cp} 
together with the results of Fig.~\ref{fig:spinodal} allow us to give a
comprehensive picture of the water anomalies and their relation to the
vapor-liquid spinodal.
The corresponding plot is presented in Figure~\ref{fig:full_scenario}.
Notice that the lines of $\kappa_T$  and $C_p$ maxima approach one to another
at high pressures and move away as the pressure decreases.
In the SF scenario both curves would only converge at 0~K.
Thus, although both the SF and LLCP scenarios are compatible with our results,
the rate of convergence of these lines seem to favor the LLCP hypothesis.
This conclusion is also supported by a comparison of Fig.~\ref{fig:full_scenario}
with the corresponding one for ST2.\cite{poole05}
The scenarios of the ST2 and TIP4P/2005 water models are completely analogous
and suggest that a critical point for TIP4P/2005 is very plausible.
In fact, we have extended the lines of $\kappa_T$ and $C_p$ maxima up to the
location of the LLCP proposed in recent papers.\cite{yagasaki14,singh16}
The extended lines provide a smooth transition from the critical point to our
simulation results for the response functions maxima.

\section{Concluding remarks}

In this work we have calculated the loci of the extrema of several
thermodynamic functions for the TIP4P/2005 water model.
In particular, the maxima and minima of the isothermal compressibility along
isobars and isobaric heat capacity along isotherms have been
evaluated and put in connection with the maxima and minima of the density
along isobars.
Our work provides for the first time a comprehensive picture of the
thermodynamic water anomalies for TIP4P/2005 and their relation to the
vapor-liquid spinodal.

The interpretation of previous simulations for the TIP4P/2005 water model in
the supercooled region has been rather controversial.%
\cite{abascal10,limmer13,overduin13,yagasaki14,overduin15,yagasaki15}
The debate has been mainly focussed on three issues: spontaneous phase
separation, finite site effects and ice coarsening.
Most of the calculations of this work correspond to the supercooled and/or the
stretched regions though we have deliberately avoided the vicinity of the
conjectured liquid-liquid region.
Our results are then beyond the current debate on the possibility of observing
spontaneous liquid-liquid phase separation.\cite{yagasaki14,overduin15,limmer15,yagasaki15}
This has allowed us to get converged results for the properties of interest
with a large but affordable computational effort.

The size of the system, 4\,000 water molecules, seems to ensure that
our calculations are free of finite size effects.
It has been reported that even larger samples could be needed for temperatures 
below the proposed LLCP.\cite{overduin15}
However, our calculations for the TMD indicate that the differences between the
results obtained with 500 and 4\,000 molecules are marginal (see Fig.~4 of
supplemental material).
Thus, at least at the thermodynamic conditions of this work, we do not observe
a significant system size dependence.

It has been argued that the phenomenon suggesting metastability of two
distinct liquid phases is actually coarsening of the ordered ice-like
phase.\cite{limmer13}
Again, the range of temperatures and pressures of this work indicate that
our results are free of the problem of ice coarsening.
In a recent study, Espinosa {\em et al.}\cite{espinosa14} have calculated the
size of the critical cluster and the nucleation rate for the crystallization
of TIP4P/2005 water as a function of the supercooling. The results for both
magnitudes indicates beyond any doubt that our simulations correspond to
a metastable liquid.
Inherent to metastability is the formation and breaking of small clusters of
the stable phase. Thus the question is not the appearance of small crystal
nuclei but whether a critical cluster may appear in the simulation.
Espinosa {\em et al.} have evaluated the supercooling required for the formation
of a single critical cluster in a simulation with a box side of 40 \AA\
(corresponding to a typical supercooled water density of about 0.94 g/cm$3$ in a
system of 2\,000 molecules) for 1 $\mu$s.
At these conditions (very similar to those of our longest simulations) they
report a 65~K supercooling. Since the melting temperature of the model is
around 250~K, the appearance of a critical cluster above 185~K is a very
unlikely event (notice that the lowest temperature of our calculations is
195~K).

Although in this work we have avoided the issue of the existence of a LLCP,
it is evident that the overall picture is consistent with both the SF and LLCP
conjectures. However, the way in which the lines of maximum $\kappa_T$ and $C_p$
approach one to another seem to indicate that they meet not too far from 
the region of calculations clearly favoring the LLCP hypothesis over the SF one.
This idea is reinforced when one observes that the scenario presented in 
Fig~\ref{fig:full_scenario} very much resembles that of ST2\cite{poole05}
for which most authors give for demonstrating the existence of a LLCP.

It is important to stress that the significance of this work goes beyond
the theoretical interpretation of simulation results.
Most of the thermodynamic states relevant to this work corresponds to the
negative pressures region where the water properties are largely
unknown.\cite{caupin14} 
Since the TIP4P/2005 water model has demonstrated to provide semiquantitative
predictions of the water properties in the supercooled and/or stretched
regions,\cite{abascal11,pallares14,pallares16} the scenario of the water
anomalies predicted by this model may provide a hint as to where the long-sought
for extrema in response functions might become accessible to experiments.

{\it Note added in proofs}: After sending the accepted version of this
work we have been aware of a paper by Lu et al.\cite{lu16} reporting
a similar study using the coarse grained mW and mTIP4P/2005 water models.

\section*{Acknowledgments}
This work has been funded by grants FIS2013-43209-P of the MEC and the Marie
Curie Integration Grant PCIG-GA-2011-303941 (ANISOKINEQ).
C.V. also acknowledges financial support from a Ram\'on y Cajal Fellowship. 
This work has been possible thanks to a CPU time allocation of the RES
(QCM-2014-3-0014, QCM-2015-1-0029 and QCM-2016-1-0036). 
We acknowledge Francesco Sciortino for valuable comments at the early stages of this work and Carlos Vega for helpful discussions.

\bibliographystyle{apsrev}

%\bibliography{./ice,add}

\section{Supplemental materials for 
"A comprehensive scenario of the thermodynamic anomalies of water using the TIP4P/2005 model"}

\subsection{Calculation of the uncertainty}

The uncertainty has been calculated using a method proposed by Hess\cite{hess02b}.
The trajectory is divided in n$_{block}$ blocks of length t$_{block}$ and the average of the density for each block is calculated. The uncertainty of the total trajectory is then calculated as the standard deviation of the values of the n$_{block}$ averages.
Notice that these uncertainties are dependent of the number of blocks (or the block length).
If t$_{block}$ is very small, consecutive blocks are strongly correlated and the standard deviation does not represent the actual uncertainty of the total trajectory.
As the block length increases, the correlation between blocks decreases so the uncertainty tends to an asymptotic value.
However, for very large block length (or, more specifically, when n$_{block}$ is low), the number of points to calculate the uncertainty is very reduced so the standard deviation shows a large statistical noise.
In fact, well known statiscal considerations indicate that n$_{block}$ should not be less than about 20.
\begin{center}
\begin{figure}[!ht]
\caption{Densities along a 400~ns trajectory for the state point at T=200~K, p=-125~MPa.}
\includegraphics*[clip,scale=0.6]{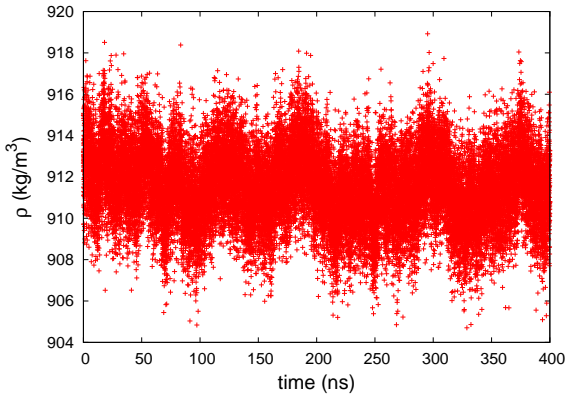}
\label{fig:dens-traject}
\end{figure}
\end{center}
\begin{center}
\begin{figure}[!ht]
\caption{Uncertainties for the system of Fig.~\ref{fig:dens-traject}.
  Symbols are the standard deviations of the averages calculated in blocks of length t$_{block}$. The line is the standard deviation evaluated using a double exponential fit for the correlation between blocks.}
\includegraphics*[clip,scale=0.6]{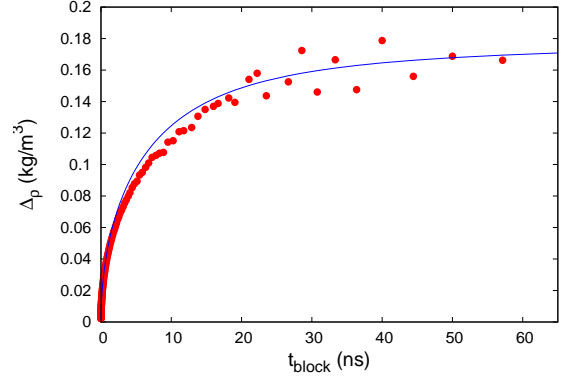}
\label{fig:errest}
\end{figure}
\end{center}
The procedure of Ref.~\onlinecite{hess02b} incorporates an important additional element, namely the calculation of the correlation between block averages and its fit to a double exponential.
In this way, the discrete nature of the correlation between blocks is smoothed by the fit and the calculated uncertainties (eventually) lead to an asymptotic curve (for long enough trajectories).
In summary, the procedure of Hess not only provides a reasonable evaluation of the uncertainty but also sheds light on the convergence of the trajectory.
An example of the application of the method is presented in Figs.~\ref{fig:errest} and \ref{fig:errestsqr}.

\begin{center}
\begin{figure}[!ht]
\caption{Block analysis of the uncertainties for the system of Fig.~\ref{fig:dens-traject} as a function of the square root of the block length.
The data correspond to the results for the full 400~ns trajectory (bottom) and for the first 100~ns (top). Notice that the number of blocks corresponding to the larger block length is only 4 in both cases. The analysis clearly indicates that a 100~ns trajectory is not long enough to obtain convergence.}
\includegraphics*[clip,scale=0.6]{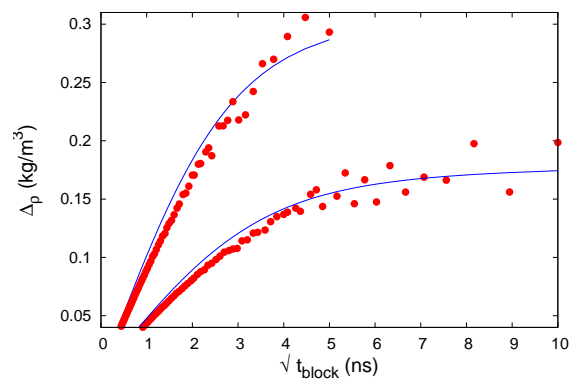}
\label{fig:errestsqr}
\end{figure}
\end{center}

\subsection{Numerical values for the density at selected isobars}
Table~\ref{table:dens} presents the numerical values of the simulation results
for the density of the TIP4P/2005 model along the isobars shown in Fig.~1 of
the main paper. 
\begin{table}[!ht]
\caption{Densities of the TIP4P/2005 model along isobars}
\begin{tabular}{ccccccc}
\hline \hline
Temperature
 & Pressure
      & Density
                & Pressure
                       & Density &Pressure& Density \\
 (K)&(MPa)&(kg/m$^3$)&(MPa)&(kg/m$^3$)&(MPa)&(kg/m$^3$) \\
\hline
 195 & 0.1 & 939.3  & -40  & 930.1  & -80  & 922.9  \\
 200 & 0.1 & 938.4  & -40  & 929.0  & -80  & 921.7  \\
 205 & 0.1 & 939.6  & -40  & 930.4  & -80  & 920.7  \\
 210 & 0.1 & 945.4  & -40  & 932.0  & -80  & 921.9  \\
 220 & 0.1 & 957.5  & -40  & 939.5  & -80  & 925.7  \\
 230 & 0.1 & 972.85 & -40  & 950.1  & -80  & 932.0  \\
 240 & 0.1 & 984.77 & -40  & 961.1  & -80  & 940.35 \\
 247 & 0.1 & 990.68 & -40  & 967.4  & -80  &   -    \\
 260 & 0.1 & 997.24 & -40  & 975.58 & -80  & 953.99 \\
 270 & 0.1 & 999.38 & -40  & 978.63 & -80  & 957.48 \\
 280 & 0.1 & 999.66 & -40  & 979.71 & -80  & 958.89 \\
 290 & 0.1 & 998.48 & -40  & 979.04 & -80  & 958.39 \\
 300 & 0.1 & 996.19 & -40  & 977.06 & -80  & 956.43 \\
 310 & 0.1 & 992.95 & -40  & 973.83 & -80  & 953.13 \\
 320 & 0.1 & 988.77 & -40  & 969.64 & -80  & 948.62 \\
 340 & 0.1 & 978.36 & -40  & 958.71 & -80  &   -    \\
 360 & 0.1 & 965.52 & -40  & 944.84 & -80  & 921.07 \\
%380 & 0.1 & 950.61 & -40  & 928.48                 \\
%400 & 0.1 & 933.79                                 \\
\hline  \hline 
\end{tabular}
\label{table:dens}
\end{table}

\subsection{Finite size effects on the TMD}
\begin{center}
\begin{figure}[!ht]
\caption{Locus of the maximum density temperatures for systems made of 500 (blue) and 4000 water molecules (red).}
\includegraphics*[clip,scale=0.6]{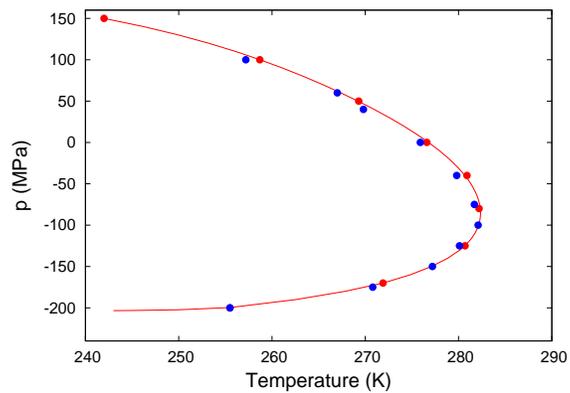}
\label{fig:TMD_500-4000}
\end{figure}
\end{center}

Fig.~\ref{fig:TMD_500-4000} shows a comparison of the TMD line calculated using
500 (data taken from Ref.~\onlinecite{pallares16}) and 4000 (this work) water
molecules.
The differences are quite small but systematic.
Since the calculations were obtained using slightly different simulation
parameters and also different version of GROMACS, it is very difficult to know
whether the small departures are a consequence of the differences in the system
size or in the simulation details.

%\subsection{Extended scenario with a LLCP}
%%
%The addition of a LLCP to the scenario obtained in this work (Figure
%\ref{fig:LLCP}) indicates that our results are fully consistent with the
%location of the proposed critical point.\cite{yagasaki14,singh16} The slope of
%the $C_p^{max}$ and $k_T^{max}$ lines need not to be modified to converge to
%the LLCP. This fact together with the similarity of Fig.~\ref{fig:LLCP} to the
%corresponding one for ST2\cite{poole05} gives a strong support for the
%conjectured existence of a second critical point in TIP4P/2005 water.
%%
%\begin{center}
%\begin{figure}[!ht]
%\caption{Comprehensive scenario of thermodynamic water anomalies for TIP4P/2005.
%In addition to the results of this work (points) we have added the proposed
%location of the LLCP.}
%\includegraphics*[clip,scale=0.6]{fig_full_scenario/full_scenario_LLCP}
%\label{fig:LLCP}
%\end{figure}
%\end{center}
%

\end{document}